# Graphene formation on SiC substrates

B.L. VanMil[1,a], R.L. Myers-Ward[1,b], J.L. Tedesco[1,c], C.R. Eddy, Jr.[1,d], G.G. Jernigan[1,e], J.C. Culbertson[1,f], P.M. Campbell[1,g], J.M. McCrate[1], S.A. Kitt[1] and D.K. Gaskill[1,h]

[1] Electronics Science and Technology Division, U.S. Naval Research Laboratory, 4555 Overlook Avenue, SW, Washington, DC 20375, U.S.A.

[a]vanmil@amethystresearch2.com, [b]rward@ccs.nrl.navy.mil, [c]tedesco@estd.nrl.navy.mil, [d]eddy@estd.nrl.navy.mil, [e]glenn.jernigan@nrl.navy.mil, [f]jim.culbertson@nrl.navy.mil, [g]paul.campbell@nrl.navy.mil, [h]kurt.gaskill@nrl.navy.mil

**Abstract.** Graphene layers were created on both C and Si faces of semi-insulating, on-axis, 4H- and 6H-SiC substrates. The process was performed under high vacuum (<$10^{-4}$ mbar) in a commercial chemical vapor deposition SiC reactor. A method for $H_2$ etching the on-axis substrates was developed to produce surface steps with heights of 0.5 nm on the Si-face and 1.0 to 1.5 nm on the C-face for each polytype. A process was developed to form graphene on the substrates immediately after $H_2$ etching and Raman spectroscopy of these samples confirmed the formation of graphene. The morphology of the graphene is described. For both faces, the underlying substrate morphology was significantly modified during graphene formation; surface steps were up to 15 nm high and the uniform step morphology was sometimes lost. Mobilities and sheet carrier concentrations derived from Hall Effect measurements on large area (16 mm square) and small area (2 and 10 μm square) samples are presented and shown to compare favorably to recent reports.

## Introduction

The predicted and measured properties of graphene have generated interest in making large area samples to test new technological applications. Since Berger *et al.*'s [1] initial report of graphene formation by the thermal desorption of Si from SiC, efforts have been underway to use this method to make large area sheets of graphene. Yet, forming large areas of uniform, electronic grade graphene by the thermal desorption process is very challenging. In this work we describe how an Aixtron VP508 SiC chemical vapor deposition (CVD) reactor using the thermal desorption method can form graphene sheets with properties similar to the best reported to-date. A two-step process is described that starts with an *in-situ* $H_2$ etch preparation of the on-axis (0001) substrate subsequently followed by the formation of graphene; the properties of the graphene are briefly described.

## Experimental

Prior investigations have shown that SiC decomposes *in vacuo* at temperatures above 1200°C and, other than residual surface C, the major components are gas phase Si and $Si_2C$ [2]; the vapor pressure of the former is estimated to be about $10^{-7}$ mbar initially rising to about $10^{-5}$ mbar at 1500°C. This suggests that a CVD reactor operating near these pressures could be able to form graphene. Preliminary work in this laboratory demonstrated the 225 mm diameter process tube of the VP508 reactor can be operated at pressures below $10^{-4}$ mbar at 1600°C using the turbo pump (Pfeiffer TMH 521) in conjunction with intermittent use of a mechanical pump (Pfeiffer Duo 2.5C) to remove $H_2$ from the cell which is not efficiently removed us-ing a turbo pump. In addition, the pumping manifold was heated to *ca*. 50°C as an aid to gas desorption. We call this the high vacuum (HV) approach to graphene.

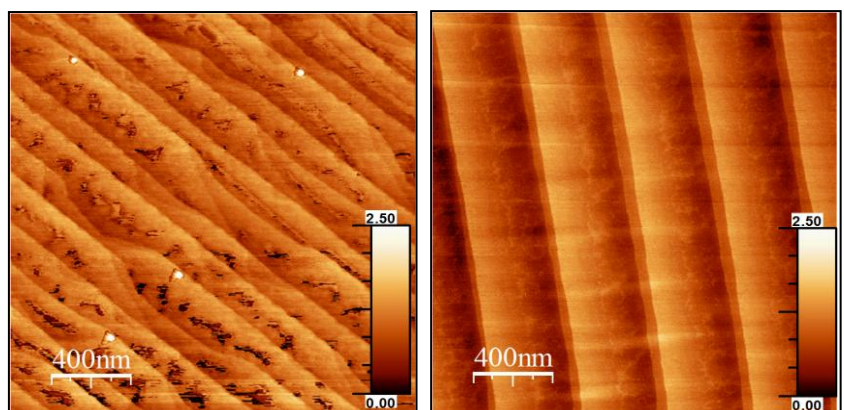

Figure 1. Atomic force map of Si-face (left) and C-face (right) 6H SiC after 5 minutes of $H_2$ etching at 1600C. The profile scales (inset) are in nanometers.

Previous $H_2$ + propane etching experiments have been described elsewhere [3] for off-axis substrates and showed the (0001)Si surface was dominated by atomic steps 0.5 nm high. This work used a similar etching formula for on-axis, semi-insulating substrates; it was found that no propane flow during $H_2$ etching produced the best surfaces. The $H_2$ etching time was about 5 minutes at 1600°C using 50 SLM at a pressure of 100 mbar and some etching occurred during temperature ramp up and down periods; the resulting total etch depth is estimated to be about 300 nm [4]. Figure 1 shows the atomic force microscopy (AFM – Digital Instruments Nanoscope IIIa, diamond-like carbon-coated tips were used) maps of Si-face and C-face 6H-SiC surfaces after $H_2$ etching. The Si-face was dominated by steps 0.50 nm high on both polytypes whereas for the C-face, steps were 1.0 nm (1.0 to 1.5 nm) high for 4H (6H) samples.

Four 16 mm x 16 mm substrates of on-axis (c-axis misorientation typically <0.5°), semi-insulating SiC substrates were prepared for each graphene experiment and consisted of both chemo-mechanically polished Si- and C-faces of the 4H (Cree) and 6H (II-VI, Inc.) polytypes. These were then $H_2$ etched as described above. To form graphene, two approaches were used. In the first, the $H_2$ flow was ramped to zero while the process tube pressure was lowered to the base pressure (~few mbar) of the process pump (Ebara A25S) and the sample was stabilized at the desired temperature. The process pressure control was then transferred to the turbo pump, and the pressure over the sample was reduced to the range ≤0.1 to $4 \times 10^{-4}$ mbar. These experiments were often limited in duration as the turbo pump could not pump the evolved $H_2$ from the cell and maintain a low pressure. Since process pressure control for the first approach was found to be less than optimal, a second approach was employed that used an Ar flush before graphene formation and the mechanical pump to maintain the pressure below $4 \times 10^{-4}$ mbar. In this approach, after the $H_2$ etch, an Ar flow (15 SLM) replaced the $H_2$ flow and the sample was brought to the desired temperature with the process tube pressure controlled by the process pump base pressure. After temperature stabilization, the process pressure control was transferred to the turbo pump and the pressure of the sample reduced to ≤$2 \times 10^{-4}$ mbar. If the base pressure exceeded $2 \times 10^{-4}$ mbar, the mechanical pump was added to the pressure control for 2 minute increments until low pressure was again achieved. For both approaches, after the graphene formation step, the sample cooled under turbo pumped vacuum.

Graphene formation was attempted at 1400, 1425, 1500, and 1600°C for durations of 10, 60, or 90 minutes. After growth, the samples were characterized by Nomarski microscopy, AFM, scanning electron microscopy (SEM), Raman spectroscopy (532 nm excitation, spot sizes were 50 μm and 2 μm) and Hall Effect measurements. Some samples were processed for additional Hall measurements using standard lithographic techniques to form a pattern of 2 and 10 μm crosses where the graphene crosses connect to Au contact pads.

## Results and Discussion

Most samples were probed with Raman spectroscopy and graphene formation was confirmed as the spectra exhibited the distinctive D, G, and 2D Raman lines [5].

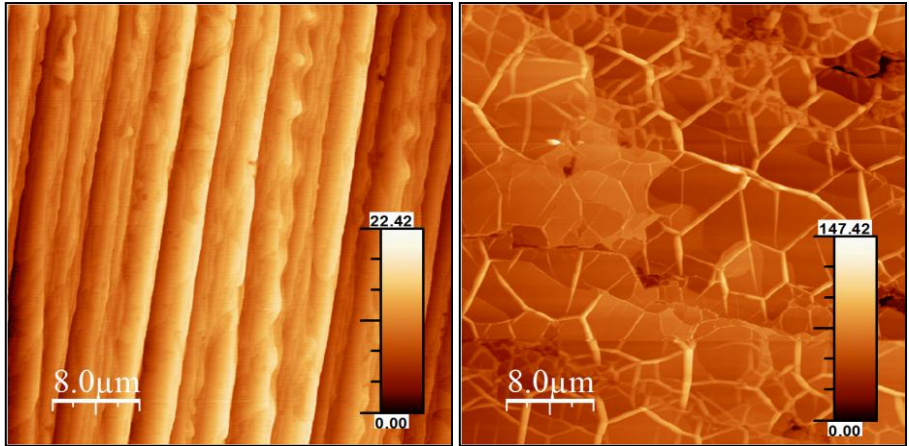

Figure 2. Atomic force map of Si-face (left) and C-face (right) 6H SiC after graphene growth. The profile scales (inset) are in nanometers.

The morphology of graphene on Si-face substrates was very different from C-face substrates. Figure 2 shows examples of both morphologies as measured by AFM. Graphene on Si-face, shown in Fig. 2 (left), was rougher than the just etched substrates as shown in Fig. 1(left). No obvious trends with substrate polytype were discerned. Removal of the graphene by the "Scotch Tape" method demonstrated that the underlying substrate morphology was nearly identical to that of the graphene. This underlying morphology usually showed step bunching with heights ranging from 1 to 15 nm depending upon growth conditions. In some cases, the steps were no longer regular and appeared to be etched irregularly. Graphene thickness measured by AFM on patterned samples ranged from about 0.5 to 8 nm, depending upon processing conditions. The error in thickness measurements is estimated to be ~1nm.

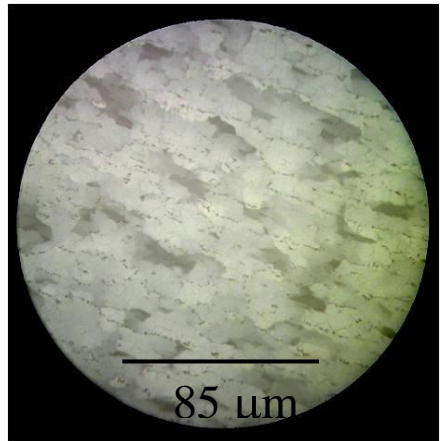

Figure 3. Nomarski micrograph of graphene on C-face 4H-SiC, same sample as in Fig. 2 (right). Note the texture across the sample.

The highest 300 K mobilities for Si-face growth recorded for whole substrate measurements were obtained for 1500 and 1600°C growths of 90 minutes duration utilizing the Ar flush; the best 300 K mobility and electron concentration were 620 $cm^2V^{-1}s^{-1}$, $8.8 \times 10^{12}$ $cm^{-2}$ (AFM thickness was 1 to 3 nm). This same sample had higher 77 K mobility values; 1560 $cm^2V^{-1}s^{-1}$ was found for an electron concentration of $4.7 \times 10^{12}$ $cm^{-2}$. In general the Hall cross samples prepared with the Ar flush showed the most uniform electrical properties. The best result was from a 10 µm cross on a 1600°C film with a thickness between 0.25 and 1 nm having a mobility of 860 $cm^2V^{-1}s^{-1}$ for an electron concentration of $1.13 \times 10^{13}$ $cm^{-2}$. These results are similar to or better than recent Hall mobility reports such as 1100 $cm^2V^{-1}s^{-1}$ (T=4 K) on a sample with an electron concentration of $3.6 \times 10^{12}$ $cm^{-2}$[1].

For the case of C-face growth, the graphene morphology was much rougher and in most cases did not clearly show the regular step morphology of the starting $H_2$ etched substrates. No obvious trends with substrate polytype were discerned. As shown in Fig. 2 (right), the graphene surface contained ridges which were not always uniform in height, ranging from 0 to 50 nm (AFM measurement), and some were estimated to have a thickness of 50 nm or less (SEM measurement). These ridges often intersected at angles that were multiples of 30 degrees but for some sample areas this was not true. It was also observed that the ridges sometimes followed underlying step edges and other times cut across steps. Removal of the graphene by the Scotch Tape method showed that the underlying sub-strate morphology was nearly identical to the graphene *except* for the ridge structure which disappeared. This underlying morphology exhibited step height variations of 1 to 15 nm. On a larger scale the morphology, shown in the Nomarski micrograph in Fig. 3 (the same sample as in Fig. 2 (right)) was not uniform. Thickness measurements using AFM on patterned graphene ranged from about 5 to 30 nm.

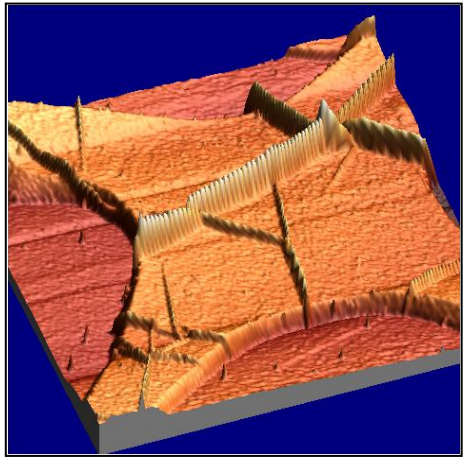

Figure 4. Atomic force microscopy map of a 2 μm cross having 300K hole mobility and concentration of 11,600 $cm^2V^{-1}s^{-1}$ and $1.54x10^{13}$ $cm^{-2}$, respectively.

The best 300 K mobilities for C-face growth recorded for whole substrate measurements were obtained for growths done at 1500°C having duration of 10 minutes using the Ar flush approach. The best 300 K mobility and associated electron concentration were 2160 $cm^2V^{-1}s^{-1}$ and $1.24x10^{13}$ $cm^{-2}$. Hall mobilities at 77 K were higher and the best was 3460 $cm^2V^{-1}s^{-1}$ for an electron concentration of $2.44x10^{13}$ $cm^{-2}$. The best 300 K mobility value was on processed samples was 18,100 $cm^2V^{-1}s^{-1}$ for a hole concentration of $2.1x10^{12}$ $cm^{-2}$ on a 10 μm cross that was 2 to 4 nm thick; Figure 4 shows the AFM measured morphology of a 2 μm cross in three dimensional relief (AFM thickness 11.7 nm). These results are similar to or better than recent Hall mobility reports such as 9500 $cm^2V^{-1}s^{-1}$ (T=180 mK) on a sample with electron concentration $3.7x10^{12}$ $cm^{-2}$ [6].

## Summary

Graphene films were synthesized in a commercial SiC epitaxial reactor using a 2-step process that combines *in-situ* $H_2$ etching of the on-axis 4H- or 6H-SiC semi-insulating substrate to remove polishing damage with a HV processing step for graphene formation. Film morphology is described and selected electrical properties by Hall Effect measurements are presented. The electrical properties are similar to or better than recent reports and thus demonstrate that high quality graphene formation in epitaxial reactors using HV is a reality.

**Acknowledgements**
This work was supported by the Office of Naval Research. BLV and RLM-W and JLT acknowledge support from the ASEE for Postdoctoral Research Fellowships. JMM and SAK acknowledge support from the Naval Research Enterprise Intern Program.

**References**

[1] C.Berger, Z.Song, T.Li, X.Li, A.Y.Ogbazghi, R.Feng, Z.Dai, A.N.Marchenkov, E. H.Conrad, P.N.First, and W.A.de Heer, J. Phys. Chem. B Vol.108 (2004) p.19912.
[2] S.S.Lilov, Cryst. Res. Technol Vol.28 (1993) p.503.
[3] K.K.Lew, B.L.VanMil, R.L.Myers-Ward, R.T.Holm, C.R.Eddy, Jr., and D.K.Gaskill, Mater. Sci. Forum Vols.556-557 (2007) p.513.
[4] B.L.VanMil, K.-K.Lew, R.L.Myers-Ward, R.T.Holm, D.K.Gaskill, C.R.Eddy, Jr., L.Wang and P.Zhao, submitted J. Cryst. Growth.
[5] E. Rollings, G.-H.Gweon, S.Y.Zhou, B.S.Mun, J.L.McChesney, B.S.Hussain, A.V.Fedorov, P.N.First, W.A.de Heer, and A.Lanzara, J. Phys. Chem. Sol. Vol.7 (2006) p.2172.
[6] C.Berger, Z.Song, X.Li, X.Wu, N.Brown, D.Maud, C.Naud, W.A.de Heer Physica Status Solidi A Vol.204 (2007) p.1746.